\documentclass[twocolumn,prb,amsmath,amssymb,floatfix]{revtex4}
\usepackage{graphicx}
\usepackage{dcolumn}
\usepackage{bm}

\begin{document}

\title{Two-chamber lattice model for thermodiffusion in polymer solutions}

\author{Jutta Luettmer-Strathmann}
\email{jutta@physics.uakron.edu}
\affiliation{%
Department of Physics, The University of Akron \\
Akron, OH 44325-4001}

\date{\today}

\begin{abstract}
When a temperature gradient is applied to a polymer solution,
the polymer typically migrates to the colder regions of the fluid
as a result of thermal diffusion (Soret effect).
However, in recent thermodiffusion experiments on
poly(ethylene-oxide) (PEO) in a mixed ethanol/water solvent 
it is observed that for some solvent compositions the polymer migrates
to the  cold side, while for other compositions it migrates to the
warm side.
In order to understand this behavior, 
we have developed a two-chamber lattice model approach to investigate
thermodiffusion in dilute polymer solutions. 
For a short polymer chain in an incompressible, one-component solvent
we obtain exact results for the partitioning of the polymer between a
warm and a cold chamber.
In order to describe mixtures of PEO, ethanol,
and water, we have extended this simple model to account for
compressibility and hydrogen bonding between PEO and water molecules. 
For this complex system, we obtain approximate results for the
composition in the warmer and cooler chambers that allow us to
calculate Soret coefficients for given temperature, pressure, and
solvent composition. 
The sign of the Soret coefficient is found to change from negative
(polymer enriched in warmer region) to positive (polymer enriched in
cooler region) as the water content of the solution is increased, 
in agreement with experimental data.
We also investigate the temperature dependence of the Soret effect and
find that a change in temperature can induce a change in the sign of
the Soret coefficient. 
We note a close relationship between the
solvent quality and the 
partitioning of the polymer between the two chambers,
which may explain why negative Soret coefficients for polymers
are so rarely observed.

\end{abstract}

\maketitle

\section{Introduction}

A temperature gradient applied to a fluid mixture generally
induces a net mass flow, which results in the formation of a
concentration gradient. This effect is known as thermodiffusion or
the Ludwig-Soret effect.\cite{de84,ha69,ty61,po62} 
The Soret coefficient $S_\mathrm{T}$ 
relates the steady state concentration gradient to
the imposed temperature gradient. 
By convention,
the Soret coefficient of component $i$
is positive if component $i$ is enriched in the cooler region.\cite{ko02} 
Thermal diffusion has long been used as an effective tool for 
separating mixtures of isotopes.\cite{po62} 
More recently, the effect has been used to characterize 
mixtures of complex fluids (see for example
Refs. \onlinecite{ko02,ru94,ko00,sh95b}).  

In liquid mixtures whose components differ widely in 
molecular mass, such as polymer solutions \cite{ru94,ko00} and colloidal
suspensions \cite{sh95b},  
it is typically the heavier component that migrates to the cold
region. 
There are, however, exceptions. 
In 1977, Giglio and Vendramini 
found a negative Soret coefficient for poly(vinyl alcohol) in
water.\cite{gi77} 
Very recently, de Gans {\it et al.} \cite{de02,de02b}
reported results of thermal diffusion forced Raleigh
scattering (TDFRS) measurements on solutions of poly(ethylene oxide)
(PEO) in mixtures  of ethanol and water. 
In pure water, PEO shows the expected migration
to the cold region of the fluid ($S_\mathrm{T}>0$). 
However, in solutions with low water
content, PEO is found to migrate to the warmer region of the
fluid ($S_\mathrm{T}<0$). 
Although 
changes in sign of the Soret coefficient have been reported for 
a number of liquid mixtures of small-molecule fluids, 
including alcohol solutions,\cite{va46,pr50,ko88,zh96b} 
the PEO/ethanol/water system 
appears to be the first polymer solution for which such a sign change
has been observed. 

Thermodiffusion in a binary fluid mixture is described by the flux of
one of the components in response to a temperature and concentration
gradient.\cite{de84} The flux is given by
\begin{equation}\label{j1}
J_1= - \rho D\nabla c_1 - \rho\, c_1 (1-c_1) D'\nabla T \, , 
\end{equation}
where $D$ is the mutual diffusion coefficient, $D'$ the thermal
diffusion coefficient of component 1, $\rho$ the total mass density,
$c_1$ the mass fraction of component 1, and $T$ is the
temperature.
Here the pressure is assumed to be constant throughout
the mixture and the flux $J_1$
describes the flow of component 1 with respect to the center of mass
of the system.\cite{de84}
Eventually, the system reaches a stationary state in which the flux
$J_1$ vanishes. Inserting $J_1=0$ into Eq.~(\ref{j1}) yields 
\begin{equation}\label{dprimed}
-\frac{1}{c_1(1-c_1)}\frac{\nabla c_1}{\nabla T}=\frac{D'}{D} .
\end{equation}
The Soret coefficient of component 1 is 
the ratio of thermal and mutual diffusion coefficients 
\begin{equation}\label{stdef}
S_\mathrm{T}=\frac{D'}{D} .
\end{equation}
More generally, we define
the Soret coefficient of component $i$ of a mixture as 
\begin{equation}\label{st}
S_\mathrm{T}=-\frac{1}{c_i(1-c_i)}\frac{d c_i}{d T} .
\end{equation}

In order to describe thermodiffusion in ternary mixtures, such as PEO
in a mixed solvent,  
concentration gradients and fluxes of two of the components
are considered.\cite{de84,la97} 
\begin{eqnarray}
\mbox{\hspace{-1em}} \label{j11}
J_1&=& - \rho D_{11}\nabla c_1 - \rho D_{12}\nabla c_2
- \rho\, c_1 (1-c_1) D_1'\nabla T\, , \\
\mbox{\hspace{-1em}} \label{j21}
J_2&=& - \rho D_{21}\nabla c_1 - \rho D_{22}\nabla c_2
- \rho\, c_2 (1-c_2) D_2'\nabla T\, , 
\end{eqnarray}
where $J_1$ and $J_2$ are the flux of component 1 and 2, respectively,
and the flux of the third component is given by $J_3=-J_1-J_2$. 
Due to Onsager's relations, only three of the 
four isothermal diffusion coefficients $D_{ij}$ are independent. 
Their relationship has a complicated form, however, since the fluxes
are defined with respect to the barycentric reference frame\cite{de84}.
The prefactors of the generalized thermal diffusion coefficients
$D_i'$ have been chosen in such a way that Eq.~(\ref{j1}) for the
binary mixture is recovered for $c_2=0$ or $c_1=0$.
In the steady state, each of the fluxes vanishes and the relationship
(\ref{dprimed}) 
between composition and temperature gradient generalizes to
\begin{eqnarray}\label{ternst}
-\frac{1}{c_1(1-c_1)}\frac{\nabla c_1}{\nabla T} &=&  \\
& & 
\mbox{\hspace{-5em}} 
\frac{c_1(1-c_1)D_{22}D_1' -c_2(1-c_2)D_{12}D_2'}
{c_1(1-c_1)(D_{11}D_{22}-D_{12}D_{21})} , \nonumber
\end{eqnarray}
and similarly for component 2. For ternary mixtures, we continue to
define the Soret coefficient of component $i$ through
Eq.~(\ref{st}).

Thermal diffusion in liquid mixtures 
is not well understood and even the sign of the Soret coefficient
cannot generally be predicted (see e.g. Refs.~\onlinecite{ko02,po62,ru94}). 
Due to the complexity of the task, attempts to extend the kinetic
gas theory\cite{ch70} of thermodiffusion to the liquid 
state have so far been unsuccessful \cite{po62,ru94}. 
Molecular
dynamics simulations (for a review see Ref. \onlinecite{ha02}) 
have become an important tool in the investigation of 
thermodiffusion in small-molecule liquids. Long computation times make
it difficult, however, 
to address thermodiffusion in polymeric systems. 

Earlier theoretical work on thermodiffusion, see e.g.\
Refs.~\onlinecite{de84,ha69,ty61,po62,wi39,wi43,wi43b,pr50b,pr50,de52}, 
made use of the ``heat of transport'' concept. 
For dilute binary solutions, the Soret coefficient of the solute
is related to the heats of transport of the two species as
follows\cite{de52}  
\begin{equation}\label{stq}
S_\mathrm{T}=\frac{Q_1^*-Q_2^*}{k_\mathrm{B}T^2} ,
\end{equation}
where $k_\mathrm{B}$ is Boltzmann's constant. The numerator 
represents the net energy flowing through the boundary of a small
volume of the mixture when the composition inside the region changes
while temperature and pressure remain constant \cite{ha69}.
Various microscopic expressions for the heat of transport have been
obtained with the aid of lattice model calculations.
Wirtz and Hiby \cite{wi39,wi43,wi43b} 
considered two contributions to the energy required for a particle to
move from one location to another, the energy to detach the
particle from its neighbors and the energy to create a
void for the particle to move into. 
Denbigh \cite{de52} considered the energy associated with particle
diffusion across a boundary to be 
a ``certain fraction'' of the change in potential
energy, which is represented by nearest-neighbor interactions in
random-mixing approximation.\cite{de52} 
This type of approach is useful in describing thermal diffusion
in regular solutions\cite{po62} but fails for more complex systems. 
Prigogine {\it et al.} \cite{pr50} found that a model based on
energies of nearest-neighbor interactions alone \cite{pr50b} led to
contradictions 
when applied to alcohol mixtures, where the Soret coefficient changes
sign\cite{va46,pr50}. They 
argued that a ``free energy'' for detaching a molecule needs to be
considered in systems with associating solvents. By assuming that 
alcohol molecules form complexes whose disruption
results in a loss of local energy and a gain in local entropy, they
were able to explain qualitatively the sign change of the Soret
coefficient for these systems.\cite{pr50} 

In this work, we investigate the Soret effect in dilute polymer
solutions with the aid of two-chamber lattice models. 
Following traditional experimental methods,\cite{de84,ha69,ty61,po62}    
we consider a system divided into two chambers of equal size 
that are maintained at slightly 
different temperatures. Particles are free to move between the
chambers, which do not otherwise interact. 
If the pressure differences between the chambers are small enough to
be neglected, the Soret coefficient can
be determined from the difference in composition of the solutions
in the two chambers.\cite{de84,ha69,ty61,po62}    
We start by investigating a polymer chain in an
incompressible, one-component solvent in Section \ref{incompressible}. 
This allows us to introduce the general strategy and 
obtain exact results for a simple model system.
Section \ref{PEOlattice}, focuses on PEO/ethanol/water mixtures. 
In Section \ref{static} we present a lattice model for the static
properties that includes the effects of compressibility and specific
interactions between PEO and water. The corresponding two-chamber
system is introduced in Section \ref{twochamber}. Results of our
calculations are presented in
Section \ref{results} and compared with experimental data, where
available. In Section \ref{discussion} we discuss the work presented
here and provide an outlook to future work.
The appendices describe the determination of system-dependent
parameters for  PEO/ethanol/water mixtures and the relation
between physical properties and model variables.

\section{Lattice model for incompressible, dilute polymer
solutions}\label{incompressible}

\subsection{Single chamber}\label{staticinc}

We consider a simple cubic lattice of $N$ sites. Since we are
interested in dilute solutions, we assume that only one polymer chain
is present. 
A chain of $N_\mathrm{c}$ beads has $N_\mathrm{c}-1$ bonds
and occupies without overlap $N_\mathrm{c}$ contiguous lattice
sites. 
For an incompressible system, all of the remaining lattice
sites are occupied individually and represent solvent molecules. 
The filling fraction of the polymer and the solvent are 
defined as $\phi_\mathrm{p}=N_\mathrm{c}/N$ and 
$\phi_\mathrm{s}=N_\mathrm{s}/N$, respectively, with 
$\phi_\mathrm{s}=1-\phi_\mathrm{p}$.
Interactions between non-bonded chain segments on
nearest-neighbor lattice sites are described by an interaction energy
$\epsilon_\mathrm{pp}$. 
Polymer-solvent and solvent-solvent interactions are also restricted
to nearest-neighbor lattice sites and described by
interaction energies $\epsilon_\mathrm{ps}$ and $\epsilon_\mathrm{ss}$, 
respectively. 

The number of chain conformations is given by the number of
self-avoiding random walks of $N_\mathrm{c}-1$ steps and can be
enumerated exactly for short chains (see e.g.\
Refs.~\onlinecite{ne92,ne92b}).   
The probability for a given chain conformation to be realized depends
on the interactions in the system.
For a chain conformation with $m$ pair contacts between polymer segments, 
the internal energy of the polymer-solvent system is given by
\begin{eqnarray}
E(m)&=&
\left(m-\left(\frac{z}{2}-1\right)N_\mathrm{c}-1\right)\Delta\epsilon
\nonumber \\ 
& &\mbox{}
+\frac{z}{2}\left(N-N_\mathrm{c}\right)\epsilon_\mathrm{ss}+
\left(\left(\frac{z}{2}-1\right)N_\mathrm{c}+1\right)\epsilon_\mathrm{pp}
\nonumber \\
& & 
\end{eqnarray}
where $z=6$ is the coordination number of the simple cubic lattice and 
$\Delta\epsilon
=\epsilon_\mathrm{pp}+\epsilon_\mathrm{ss}-2\epsilon_\mathrm{ps}$ 
is the net interaction energy between polymer segments.
The canonical partition function of the system is given by
\begin{equation}\label{partinc}
Z_\mathrm{pol}(T)=N\sum_m c(m) \exp(-\beta E(m)) ,
\end{equation}
where $c(m)$ is the number of chain conformations with $m$ polymer-polymer 
contacts and $\beta=1/k_\mathrm{B}T$, where $T$ is the temperature and 
$k_\mathrm{B}$ is Boltzmann's constant. The internal energy for a
given temperature is calculated in the usual way 
\begin{eqnarray}\label{upol}
U_\mathrm{pol} &=&
N\sum_m c(m) E(m)\exp(-\beta E(m))/Z_\mathrm{pol} \\
&=&
k_\mathrm{B}T^2\frac{\partial\ln(Z_\mathrm{pol})}
{\partial T} .
\end{eqnarray}

\begin{figure}
\vspace*{-0.5cm}
\includegraphics*[width=3.3in]{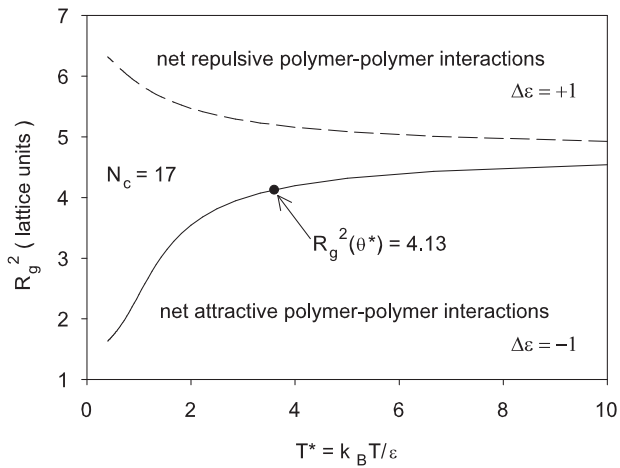}
\caption{\label{fig1}
Radius of gyration squared of chains of length $N_\mathrm{c}=17$ as a
function of reduced temperature for net attractive
($\Delta\epsilon<0$) and net repulsive ($\Delta\epsilon>0$)
polymer-polymer interactions.} 
\end{figure}
In a preliminary step, we 
generated all single chain conformations of chains up to length
$N_\mathrm{c}=17$ and determined the number $m$ of polymer-polymer
contacts for each conformation as well as the radius of gyration. 
The radius of gyration for a given temperature is calculated from
\cite{ne92,ne92b}
\begin{equation}
\langle R_g^2(T) \rangle = N \sum_m c(m) 
\bar{R_g^2}(m)\exp(-\beta E(m))/Z_\mathrm{pol} ,
\end{equation}
where $\bar{R_g^2}(m)$ is the average radius of gyration squared for 
conformations with $m$ contacts. 
For the incompressible system, we express all interaction energies in 
terms of a positive energy unit $\epsilon$, which also sets the
temperature scale; the reduced temperature $T^*$ is given by 
$T^*=k_\mathrm{B}T/\epsilon$.
In Fig.~\ref{fig1} we present chain dimensions
for chains of length $N_\mathrm{c}=17$ as a function of reduced
temperature for the case of 
net attraction ($\Delta\epsilon<0$) and net repulsion ($\Delta\epsilon>0$) 
between the polymer segments. 
As expected, the chain expands with increasing 
temperature for the case $\Delta\epsilon<0$, 
while the chain decreases in size with increasing 
temperature for the case $\Delta\epsilon>0$. 

For polymers in dilute solutions, the chain dimensions are an
indicator for solvent quality \cite{do96};  
the better the solvent the larger the chain dimensions. 
For very long chains with net attractive interactions between the
polymer segments, a collapse transition is observed at the $\theta$
temperature of the polymer solution \cite{do96}. 
Monte Carlo simulations \cite{me90} yield an estimate of
$\theta^*=3.60\pm 0.05$ 
for the reduced $\theta$ temperature of an infinitely long isolated 
chain. From our exact enumeration results we determined the dimensions
of an isolated chain of length $N_\mathrm{c}=17$ at the $\theta$ temperature 
of the infinite chain and found $R_g^2(\theta^*)=4.13$ in units
of lattice constant squared.

\subsection{Two-chamber system}\label{twochamberinc}

\begin{figure}
\includegraphics*[width=3.3in]{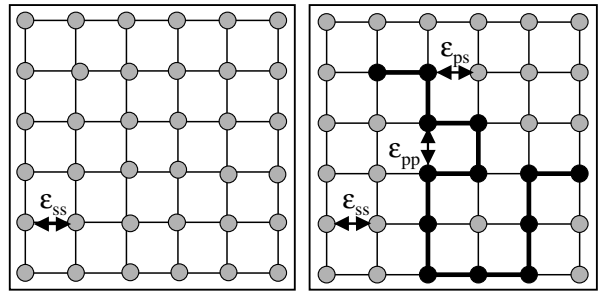}%
\caption{\label{fig2} 
Two-dimensional illustration of the two-chamber lattice model for the
incompressible case. The right chamber contains the polymer chain, 
where the black circles and heavy lines indicate the beads and bonds
of the polymer chain while the grey circles represent solvent sites. 
}
\end{figure}

In order to investigate thermodiffusion, 
consider a single-chain system that is divided into two chambers, A
and B, with slightly different temperatures,
$T^\mathrm{A}>T^\mathrm{B}$, see Fig.~\ref{fig2}. 
Under the assumption that the chambers
are non-interacting, the partition function of the whole system is
the product of the partition functions of the individual chambers, 
$Z^\mathrm{A}Z^\mathrm{B}$. The chambers are represented by lattices
with $N_a$, $a\in\{A,B\}$ sites.
The canonical partition function of the chamber that contains the polymer 
is given by Eq.~(\ref{partinc}), the canonical partition function for the 
chamber without polymer consists of a single term
\begin{equation}\label{partnopinc}
Z_\mathrm{nop}(T_a)=\exp(-\beta_a E_\mathrm{nop}) 
\end{equation}
with $a\in\{A,B\}$ and
$E_\mathrm{nop}=\frac{z}{2}N_a\epsilon_\mathrm{ss}$ .
The internal energy in this case is independent of temperature,
$U_\mathrm{nop}=E_\mathrm{nop}$.
If the polymer chain is allowed to be in either chamber, the sum of
states for the system is given by
\begin{equation}\label{suminc}
Q=Z_\mathrm{pol}(T_\mathrm{A})Z_\mathrm{nop}(T_\mathrm{B})+
Z_\mathrm{pol}(T_\mathrm{B})Z_\mathrm{nop}(T_\mathrm{A}) .
\end{equation}
Accordingly, the probability $q_\mathrm{A}$ to find the polymer in the
(warmer) chamber A can be calculated from
\begin{equation}\label{qainc}
q_\mathrm{A}=Z_\mathrm{pol}(T_\mathrm{A})Z_\mathrm{nop}(T_\mathrm{B})/Q .
\end{equation}
$q_\mathrm{B}=1-q_\mathrm{A}$ is the probability to find
the chain in the cooler chamber B.

For small temperature differences 
$\delta T=T_\mathrm{A}-T_\mathrm{B}$ we set
$T_\mathrm{B}=T$, $T_\mathrm{A}=T+\delta T$ and 
rewrite Eq.~(\ref{qainc}) as 
\begin{equation}\label{qainc2}
\frac{1}{q_\mathrm{A}}=1+
\frac{
Z_\mathrm{pol}(T)Z_\mathrm{nop}(T+\delta T)
}{
Z_\mathrm{pol}(T+\delta T)Z_\mathrm{nop}(T)} .
\end{equation}
Expanding the partition functions according to
\begin{equation}\label{expandlnz}
\ln(Z(T+\delta T))=\ln(Z(T))+\beta U(T)\delta T/T ,
\end{equation}
where $\beta=1/k_\mathrm{B} T$, we find for the probability
\begin{equation}\label{qaincfin}
\frac{1}{q_\mathrm{A}}=1+
\exp\left(
\beta\left(U_\mathrm{nop}-U_\mathrm{pol}\right)\frac{\delta T}{T}\right) .
\end{equation}
We employ temperature differences $\delta T/T=10^{-4}$ in our calculations and 
find the results from Eqs.~(\ref{qainc}) and (\ref{qaincfin}) to be
indistinguishable. 
In order to compare with the theories based on the heat of transport, 
\cite{de84,ha69,ty61,po62,wi39,wi43,wi43b,pr50b,pr50,de52}
we expand the expression for the probability 
$q_\mathrm{A}-1/2$ from Eq.~(\ref{qaincfin}) 
in powers of $\delta T/T$ and obtain to first order
\begin{equation}\label{qaincexp}
q_\mathrm{A}-\frac{1}{2} \simeq -\frac{1}{4}\beta
\left(U_\mathrm{nop}-U_\mathrm{pol}\right)\frac{\delta T}{T} .
\end{equation}
For this simple model, the probability to find the polymer in the
warmer chamber A is larger than 1/2 when the internal energy
of the chamber without polymer is smaller than the internal energy
of the chamber with polymer at temperature $T$. 
Hence, the difference in internal energy between two chambers, one
with and one without polymer, 
at the same temperature determines the probability to find the
polymer in the warmer of two chambers.

The Soret coefficient ( Eq.~(\ref{st}) )
can be calculated from the filling fraction as well 
as from the mass fraction, since $\partial c_1/\partial \phi_1=
c_1c_2/\phi_1\phi_2$. 
Setting $\phi_1=\phi_\mathrm{p}$, we obtain in the dilute limit
\begin{equation}\label{stdilute}
S_\mathrm{T} \simeq -\frac{1}{\phi_\mathrm{p}}\frac{d\phi_\mathrm{p}}{dT}
\simeq -\frac{1}{\phi_\mathrm{p}}
\frac{(\phi_\mathrm{pA} - \phi_\mathrm{pB})}
{(T_\mathrm{A}-T_\mathrm{B})} ,
\end{equation}
where the filling fractions in the chambers are calculated from the 
probability $q_\mathrm{A}$ 
\begin{equation}
\phi_\mathrm{pA}=q_AN_\mathrm{c}/N_\mathrm{A} , \mbox{\hspace{2em}}
\phi_\mathrm{pB}=(1-q_A)N_\mathrm{c}/N_\mathrm{B} .
\end{equation}
Typically we assume that the chambers are equal in size and set
$N_\mathrm{A}=N_\mathrm{B}=N/2$, which yields for the Soret coefficient 
\begin{equation}\label{stqdel}
S_\mathrm{T}=-4\frac{q_\mathrm{A}-\frac{1}{2}}{\delta T} .
\end{equation}

\begin{figure}
\vspace*{-0.5cm}
\includegraphics*[width=3.3in]{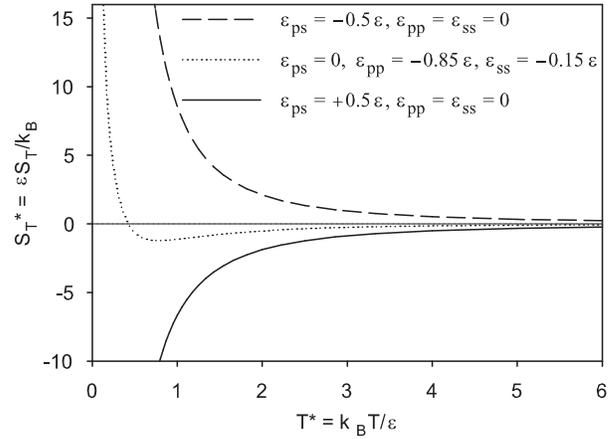}
\caption{\label{fig3}
Soret coefficients of an incompressible dilute polymer solution as a
function of temperature. The curves represent exact enumeration
results for chains of length $N_\mathrm{c}=17$, where the Soret
coefficient has been calculated from  
Eq.~(\protect{\ref{stqdel}}). 
The interaction parameters are as indicated in the figure and
correspond to net repulsive interactions 
($\Delta\epsilon=+1$) for the dashed curve and net attractive
interactions  
($\Delta\epsilon=-1$) for the dotted and solid curves,
respectively. }
\end{figure}

With the aid of Eq.~(\ref{qaincexp}) for the probability, the Soret
coefficient becomes 
\begin{equation}\label{stincapp}
S_\mathrm{T}=\frac{U_\mathrm{nop}-U_\mathrm{pol}}{k_\mathrm{B}T^2} .
\end{equation}
In agreement with the definition of the Soret coefficient, this
expression for $S_\mathrm{T}$ is independent of the temperature
difference $\delta T$. It is also independent of the polymer
concentration, in agreement with experimental results for dilute
polymer solutions (cf.\ Ref.~\onlinecite{ra02}). 
Comparing Eq.~(\ref{stincapp}) 
with the heat of transport expression for dilute solutions in 
Eq.~(\ref{stq}), we see that the 
internal energy difference $U_\mathrm{nop}-U_\mathrm{pol}$ 
takes the place of the heat of transport difference 
$Q_1^*-Q_2^*$.  

Results for the Soret coefficient calculated
from Eqs.~(\ref{qaincfin}) and (\ref{stqdel})
are presented in Fig.~\ref{fig3}. 
We note first that both positive and negative 
Soret coefficients are realized and that $S_\mathrm{T}$ may 
change sign as a function of temperature. 
For the simple model discussed in this
section, the internal energy $U_\mathrm{nop}$ of a solvent-filled
lattice is independent of temperature. 
If $U_\mathrm{pol}$ was also independent of temperature, then 
$S_\mathrm{T}$ would vary as $1/T^2$, according to
Eq.~(\ref{stincapp}). The deviation of the actual calculated Soret
coefficients from this $1/T^2$ dependence, most notably the sign
change of $S_\mathrm{T}$, is due to the temperature
variation of the internal energy $U_\mathrm{pol}$ of
the polymer system. 
A random mixing approximation for
polymer-polymer and polymer-solvent contacts 
yields a temperature-independent internal energy
$U_\mathrm{pol}$. Hence, in order to 
calculate Soret coefficients of polymer solutions, one has to go
beyond the random mixing approximation, as we do in this work.

A comparison of the results for the chain dimensions in Fig.~\ref{fig1} with
the results for the Soret coefficients represented by the dashed and
solid curves in Fig.~\ref{fig3} shows that the Soret coefficient decreases 
as the solvent quality increases. 
We find this trend confirmed when we consider the more
complex system below. 
However, the Soret coefficients represented by the dotted curve show
that this need not be the case. The chain dimensions for this case are
the same as for the case represented by the solid line, since they
correspond to the same net interaction energy $\Delta\epsilon=-1$. 
It was recognized early \cite{pr50b,de52} that, 
in contrast to thermodynamic and configurational properties, 
thermodiffusion is not determined by the net interaction energy
$\Delta\epsilon$ but by all three interaction parameters. In fact, the
simple heat of transfer models \cite{wi43b,de52,pr50b} are most
successful for regular solutions, where $\Delta\epsilon=0$.

\section{Lattice model for PEO in ethanol/water mixtures}\label{PEOlattice}

Solutions of high molecular weight poly(ethylene oxide) (PEO) in
ethanol and water have interesting properties. 
Hydrogen bonding between PEO and water molecules plays an 
important role in aqueous solutions of PEO 
(see e.g.\ Refs.~\onlinecite{je96,do02}). 
Water is a good solvent for PEO at standard temperature and 
pressure. However, the solvent quality decreases with temperature and a 
miscibility gap opens above a lower critical solution
temperature.\cite{do02} 
Ethanol, on the other hand, is a poor solvent for PEO at room
temperature but the solubility increases with temperature \cite{de02}. 
In mixtures of ethanol and water at standard temperature and pressure,
the water content determines the solubility of PEO.
For the molecular weight considered in this work, the transition
between poor and good solvent condition appears between 
a water content of 5\% and 10\% by weight \cite{de02,de02b}.
Light scattering experiments \cite{de02} show that the PEO chains
expand with increasing water content, indicating that the addition of
water improves the solvent quality.

\subsection{Static properties}\label{static}

In order to describe dilute solutions of poly(ethylene oxide) (PEO) in
mixtures of ethanol and water, we extend the lattice model introduced
above. 
At a given temperature, pressure, and composition, the single chamber
system is represented by a simple cubic lattice with 
$N$ sites, of which $N_\mathrm{c}$, $N_\mathrm{s}$, 
and $N_\mathrm{w}$
are occupied by the polymer (PEO), the first solvent
(ethanol), and the second solvent (water), respectively. In order to
account for compressibility, we allow sites to be unoccupied 
so that $N=N_\mathrm{c}+N_\mathrm{s}+N_\mathrm{w}+N_\mathrm{v}$, where 
$N_\mathrm{v}$ is the number of voids. The total volume of the lattice is
$V=v_0N$, where $v_0$ is the volume of one elementary cube.

Interactions between occupied nearest neighbor sites are described by
interaction energies $\epsilon_{ij}$, where the subscripts indicate
the occupants of the sites ($\mathrm{p}$ for polymer, $\mathrm{s}$ 
and $\mathrm{w}$ for the
solvents; voids are assumed to have zero interaction energies).
In aqueous solutions, hydrogen bonding between PEO and water plays an
important role (cf.\ Ref.~\onlinecite{je96}). 
In order to account for these specific interactions,
we introduce an orientational degree of
freedom in the description of water. Each elementary cube occupied by
water is assumed to have one special face. If this face is exposed to
a polymer segment, the interaction energy is 
$\epsilon_\mathrm{pw;s}$ (strongly attractive) otherwise
$\epsilon_\mathrm{pw;n}$ (non-specific).

Because of the complexity of the system, we can no longer evaluate the
canonical partition function exactly. 
In the following, we employ the exact enumeration results for the
chains of length $N_\mathrm{c}=17$ and consider explicitly 
contacts that involve the polymer but we employ a random mixing
approximation for all other contacts. 
A chain conformation with $m$ pair contacts has
$n_\mathrm{n}=4N_\mathrm{c}+2-2m$ nearest neighbor (nn) sites, which
are occupied by $n_i$, $i\in\{\mathrm{s,w,v}\}$ solvent particles and
voids. 
With the aid of the random mixing approximation for all but the polymer
contacts, the canonical partition function of the system can be
written as  
\begin{eqnarray} \label{Z1} 
&& \mbox{\hspace{-1.5em}}Z_\mathrm{pol}(N,T,N_\mathrm{w},N_\mathrm{s}) =\\
&&N\sum_m c(m) {\sum_{[n_\mathrm{w}]}} 
6^{N_\mathrm{w}-n_\mathrm{w}} 
\left( \begin{array}{c} n_\mathrm{n} \\ n_\mathrm{w} \end{array} \right) 
\left(\begin{array}{c} N-n_\mathrm{n}-N_\mathrm{c} 
\\ N_\mathrm{w}-n_\mathrm{w} \end{array} \right) 
\nonumber \\
&& 
\times 
{\sum_{[n_\mathrm{s}]}} 
\left(\begin{array}{c} n_\mathrm{n}-n_\mathrm{w} \\ n_\mathrm{s} \end{array} \right)
\left(\begin{array}{c}
N-n_\mathrm{n}-N_\mathrm{c}-(N_\mathrm{w}-n_\mathrm{w})  
\\ N_\mathrm{s}-n_\mathrm{s} \end{array} \right) \nonumber \\
&& \times
e^{-\beta(m\epsilon_\mathrm{pp}+n_\mathrm{s}\epsilon_\mathrm{ps})}
\left(5e^{-\beta\epsilon_\mathrm{pw;n}}+
e^{-\beta\epsilon_\mathrm{pw;s}}\right)^{n_\mathrm{w}} 
e^{-\beta E_\mathrm{r}} \: , \nonumber 
\end{eqnarray} 
where, as before, $c(m)$ is the number of chain conformations with $m$
polymer-polymer contacts. 
The square brackets around the summation indices indicate that the
summation is performed consistent 
with the available nearest neighbor sites and the 
total filling of the lattice. The energy $E_\mathrm{r}$
denotes the contribution to the total energy due to 
solvent-solvent interactions evaluated in random mixing
approximation \cite{la00}, cf.\ Eq.~(\ref{Eempty}) below. 

\begin{figure}
\vspace*{-0.3cm}
\includegraphics*[width=3.3in]{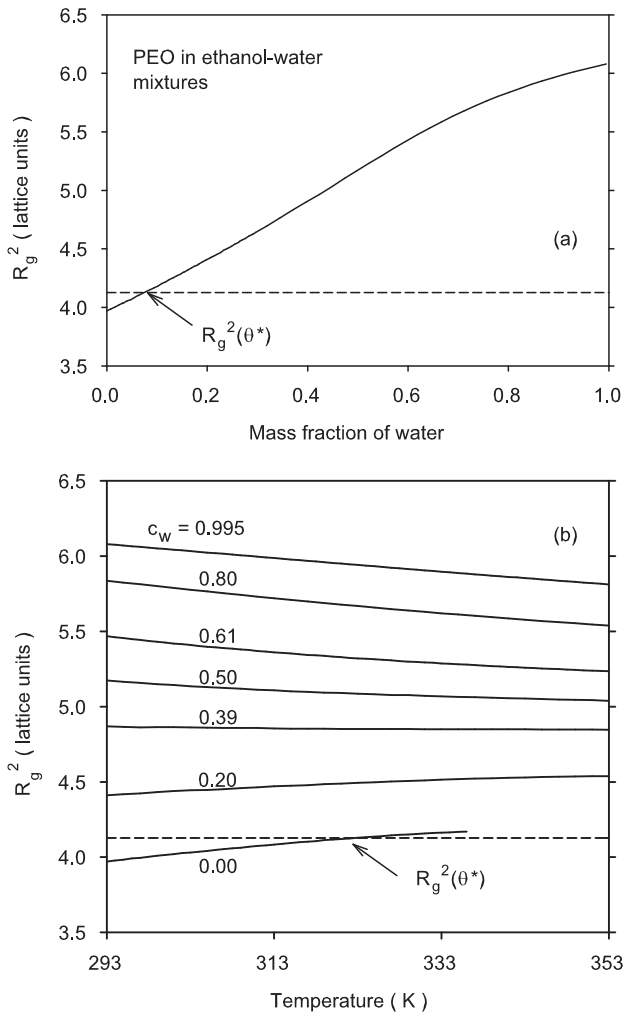}%
\caption{\label{fig4}
Radius of gyration squared, $R_\mathrm{g}^2$, as calculated from 
Eq.~(\protect{\ref{Rgsq}}). 
Panel (a) shows chain dimensions as a 
function of solvent composition at temperature $T=293$~K, pressure 
$P\approx 0.1$~MPa, and a PEO concentration of 5~g/L. 
Panel (b) shows the temperature variation of the chain dimensions
when the solutions are heated at constant pressure.
The dashed line indicates the chain dimensions,
$R_\mathrm{g}^2(\theta^*)$, of the isolated chain 
at the $\theta$ temperature of the infinite
chain, as discussed in Section~\protect{\ref{staticinc}}.
}
\end{figure}

The pressure of the system is calculated from
\begin{equation}\label{pres}
P=\frac{k_\mathrm{B} T}{v_0} 
\left(\frac{\partial \ln{Z_\mathrm{pol}}}{\partial
N}\right)_{N_\mathrm{s},N_\mathrm{w},N_\mathrm{c}} .
\end{equation}
By performing partial summations over the terms in 
Eq.~(\ref{Z1}) the probabilities for specific sets of states
can be determined. 
If we write the partition function as 
\begin{equation}\label{partialZ}
Z_\mathrm{pol} \equiv 
\sum_{m,[n_\mathrm{w}],[n_\mathrm{s}]} 
Z_{m,n_\mathrm{w},n_\mathrm{s}} \: ,
\end{equation} 
the average radius of gyration $\langle R_\mathrm{g}^2\rangle$ 
is given by
\begin{equation}\label{Rgsq}
\langle R_\mathrm{g}^2\rangle=
Z_\mathrm{pol}^{-1}
\sum_{m,[n_\mathrm{w}],[n_\mathrm{s}]} 
\bar{R}_\mathrm{g}^2(m)Z_{m,n_\mathrm{w},n_\mathrm{s}} \: .
\end{equation}

In Appendix~\ref{appa}, we discuss how the system-dependent parameters of
the lattice model were determined. In Appendix~\ref{appb} we describe
how lattice size and occupation numbers for given temperature,
pressure and composition of the mixture are calculated. 
Since chain dimensions are an indicator for solvent quality, we
employed calculated $R_\mathrm{g}^2$ values to 
obtain estimates for the mixed interaction parameters.  
In Fig.~\ref{fig4} we present graphs
 for the chain dimensions calculated with the aid of the
system-dependent parameters presented in Table~\ref{tabsys}.

\subsection{Two-chamber system}\label{twochamber}

As in section \ref{twochamberinc}, we consider a system of two chambers 
with slightly different temperatures and determine the probability
to find the polymer in the warmer of the two chambers.
Assuming again that the chambers are non-interacting, the 
canonical partition function of the system 
for a given occupation of the chambers is the product of 
the individual partition functions.

To find the required canonical partition function of a chamber
without polymer, consider a lattice of $N$ sites 
occupied by the two types of solvent and voids, 
$N=N_\mathrm{s}+N_\mathrm{w}+N_\mathrm{v}$. 
Denoting the filling fractions of ethanol and water 
by $\phi_\mathrm{s}=N_\mathrm{s}/N$ and
$\phi_\mathrm{w}=N_\mathrm{w}/N$  
and assuming random mixing, the internal energy is given by \cite{la00} 
\begin{equation}\label{Eempty}
E_\mathrm{nop}=\frac{z}{2}N\left(\epsilon_\mathrm{ss}\phi_\mathrm{s}^2+
\epsilon_\mathrm{ww}\phi_\mathrm{w}^2+
2\epsilon_\mathrm{ws}\phi_\mathrm{s}\phi_\mathrm{w}\right).
\end{equation}
Accordingly, the canonical partition function of the lattice without polymer
takes the form
\begin{eqnarray}\label{zempty}
& & \mbox{\hspace{-3em}} Z_\mathrm{nop}(N,T,N_\mathrm{s},N_\mathrm{w})
\nonumber \\
& &=6^{N_\mathrm{w}}
\left(\begin{array}{c} N \\ N_\mathrm{w} \end{array} \right) 
\left(\begin{array}{c} N-N_\mathrm{w} \\ N_\mathrm{s} \end{array}
\right) e^{-\beta E_\mathrm{nop}} .
\end{eqnarray}

To ease notation for the two-chamber sum of states, we define the
canonical partition for a single chamber as 
\begin{eqnarray}\label{zgen}
& & \mbox{\hspace{-3em}}
Z(N,T,N_\mathrm{w},N_\mathrm{s},N_\mathrm{p})  \nonumber \\
& & 
=\left\{
\begin{array}{rcl}
Z_\mathrm{pol}(N,T,N_\mathrm{s},N_\mathrm{w}) & \mbox{ for } &
N_\mathrm{p}=1 \\
Z_\mathrm{nop}(N,T,N_\mathrm{s},N_\mathrm{w}) & \mbox{ for } &
N_\mathrm{p}=0 
\end{array} \right.
\end{eqnarray}
where $N_\mathrm{p}\in\{0,1\}$ is the number of polymer chains in the
chamber. 

For the mixed solvent system considered here, the concentration of
the two solvents will generally be different in the warm and cold
regions of a fluid. 
Since we have no a priori information about concentrations (or
chemical potentials) of the solvents, we consider all 
distributions of particles consistent with fixed total particle
numbers. The sum of states is then given by
\begin{eqnarray}\label{sumofstates}
Q &=& 
\sum_{N_\mathrm{p,A}=0}^{1}
\sum_{[N_\mathrm{w,A}]}\sum_{[N_\mathrm{s,A}]}  
Z(N_\mathrm{A},T_\mathrm{A},N_\mathrm{w,A},N_\mathrm{s,A},N_\mathrm{p,A}) 
\nonumber \\ &&
\mbox{\hspace{-1.5em}} \times 
Z(N-N_\mathrm{A},T_\mathrm{B},N_\mathrm{w}-N_\mathrm{w,A},
N_\mathrm{s}-N_\mathrm{s,A},1-N_\mathrm{p,A}) , \nonumber \\ 
\end{eqnarray}
where, as before, square brackets indicate summations consistent with
the total numbers of particles and lattice sites. Also as before, 
chamber A is considered the warmer chamber
so that $\delta T=T_\mathrm{A}-T_\mathrm{B}>0$, and equal-sized
chambers are used, $N_\mathrm{A}=N_\mathrm{B}=N/2$. 
As we are performing the calculation of the terms in the sum of
states, we monitor for each chamber the composition and the pressure
of the mixtures. This allows us to calculate the average quantities
for each chamber by performing weighted sums. For example, the average
mass fraction of component $i$ in chamber A, is calculated from 
\begin{eqnarray}\label{calcmassfrac}
c_{i,\mathrm{A}}&=&\frac{1}{Q}
\sum_{N_\mathrm{p,A}=0}^{1}
\sum_{[N_\mathrm{w,A}]}\sum_{[N_\mathrm{s,A}]} 
c_i(N_\mathrm{w,A},N_\mathrm{s,A},N_\mathrm{p,A}) \\ 
& & \mbox{\hspace{-1.7em}} \times 
Z(N_\mathrm{A},T_\mathrm{A},N_\mathrm{w,A},N_\mathrm{s,A},N_\mathrm{p,A}) 
\nonumber \\ &&
\mbox{\hspace{-1.7em}} \times 
Z(N-N_\mathrm{A},T_\mathrm{B},N_\mathrm{w}-N_\mathrm{w,A},
N_\mathrm{s}-N_\mathrm{s,A},1-N_\mathrm{p,A}) , \nonumber 
\end{eqnarray}
where $c_i$ is the mass fraction of component $i$ calculated 
according to Eq.~(\ref{massfrac}). 

The probability to find the polymer in chamber A in this model is 
given by
\begin{eqnarray}\label{qacomp}
q_\mathrm{A}&=&\frac{1}{Q} 
\sum_{[N_\mathrm{w,A}]}\sum_{[N_\mathrm{s,A}]} 
Z_\mathrm{pol}(N_\mathrm{A},T_\mathrm{A},N_\mathrm{w,A},N_\mathrm{s,A}) 
 \\ &&
\mbox{\hspace{-1.7em}} \times 
Z_\mathrm{nop}(N-N_\mathrm{A},T_\mathrm{B},N_\mathrm{w}-N_\mathrm{w,A},
N_\mathrm{s}-N_\mathrm{s,A}) . \nonumber 
\end{eqnarray}
As in the case of the incompressible system, this probability is
related to an internal energy difference of two chambers at the same
temperature: 
\begin{equation}\label{qacompexp}
q_\mathrm{A}-\frac{1}{2} \simeq -\frac{1}{4}
\frac{
\langle U_\mathrm{nop}\rangle - \langle U_\mathrm{pol}\rangle}{k_\mathrm{B}T}
\frac{\delta T}{T} .
\end{equation}
The angular brackets indicate an average 
over all configurations of particles in two chambers at the same
temperature $T$, where the polymer is confined to one of the
chambers. 
$U_\mathrm{pol}$ and $U_\mathrm{nop}$ are the 
average internal energies of the chamber with and without polymer
at fixed composition.
For the small temperature differences $\delta
T=10^{-4}$~K, corresponding to $\delta T/T\simeq 3\times 10^{-7}$, 
employed in our calculations, values for the
probability $q_\mathrm{A}$ calculated from Eqs~(\ref{qacomp}) and
(\ref{qacompexp}) agree to more than five digits. 
As in the incompressible case, the excess probability
$q_\mathrm{A}-1/2$ is proportional to the temperature difference and
independent of the polymer concentration for dilute solutions.
Relations similar to Eq.~(\ref{qacompexp}) can be derived for the
probability of finding a given number of solvent particles in the
warmer chamber A. They show that the difference between the average
numbers of particles in chambers A and B is proportional to $\delta
T$.

\section{Results}\label{results}

\begin{figure}
\vspace*{-0.3cm}
\includegraphics*[width=3.3in]{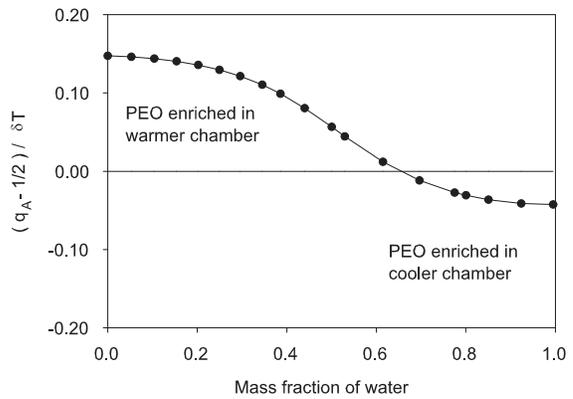}%
\caption{\label{fig5}
Excess probability 
to find the PEO chain in the warmer chamber A
as a function of water content of the solution. 
Calculations were performed for the compositions indicated by the
symbols; the symbol size is larger than the uncertainty due to
discrete occupation number discussed in Appendix~\protect{\ref{appb}}. }
\end{figure}

\begin{figure}
\vspace*{-0.3cm}
\includegraphics*[width=3.3in]{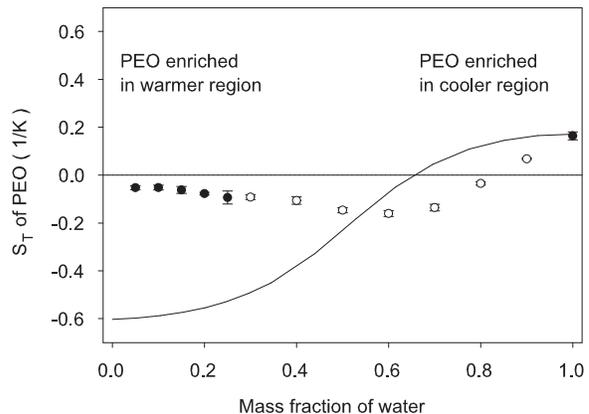}%
\caption{\label{fig6}
Soret coefficient of PEO in mixtures of ethanol and water at 
room temperature and atmospheric pressure. The filled symbols represent
experimental data by de Gans {\it et al.}\protect{\cite{de02}}, the open 
symbols represent preliminary experimental results by 
Kita {\it et al.}\protect{\cite{ki03}}, the line is the result of our 
lattice model calculations.}
\end{figure}

\begin{figure}
\vspace*{-0.3cm}
\includegraphics*[width=3.3in]{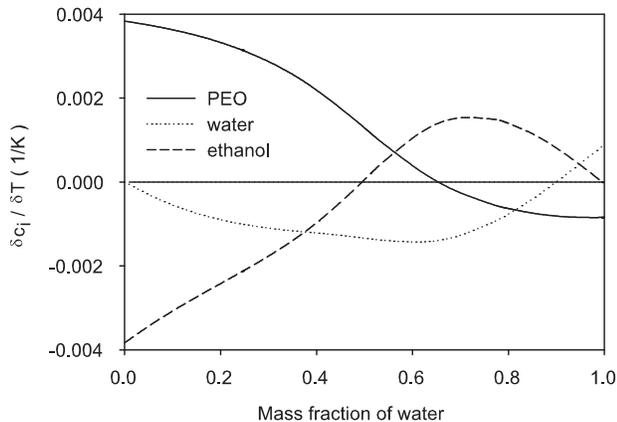}%
\caption{\label{fig7}
Composition differences $\delta c_i = c_{i,\mathrm{A}}-c_{i,\mathrm{B}}$
between chambers A and B 
for the three components (PEO, ethanol, water) as a function of 
water content of the solution. 
The compositions of the chambers were calculated from 
Eq.~\protect{\ref{calcmassfrac}} and divided 
by the temperature difference $\delta
T=T_\mathrm{A}-T_\mathrm{B}=10^{-4}$~K. 
}
\end{figure}

We have applied our lattice model to solutions of PEO in ethanol/water
mixtures under a variety of conditions.  
As noted above, calculated values for composition
differences in the two chambers are expected to be proportional to the
temperature difference $\delta T=T_\mathrm{A}-T_\mathrm{B}$ 
for sufficiently small values of $\delta T$. 
We find this to be the case for a large range of $\delta T$ values and
used $\delta T=10^{-4}$~K throughout the calculations.
Furthermore, since the probability to find the polymer in the warmer
chamber is independent of polymer concentration, we may choose the
PEO content of the solution. As discussed in Appendix~\ref{appb}, a
low polymer concentration corresponds to a large lattice size, which
reduces the uncertainty due to integer occupation numbers but
increases the computation time. 
We found that the PEO concentration of 5g/L used in the experiments
represents a satisfactory compromise.
The results presented in Figs.~\ref{fig5}--\ref{fig7} 
pertain to room temperature (293 K),
atmospheric pressure (0.1 MPa), and a PEO concentration of
5~g/L. 

In Fig.~\ref{fig5} we present values for the excess probability
$q_\mathrm{A}-1/2$ as calculated from Eq.~(\ref{qacomp}). 
In Fig.~\ref{fig6} we present the corresponding 
values for the Soret coefficient of PEO calculated according to
Eqs.~(\ref{calcmassfrac}) and (\ref{st}). 
For comparison, we include experimental data by Wiegand and 
coworkers \cite{de02,ki03}.
Both experiment and theory show a change in sign of the Soret
coefficient as the water content of the solution is increased. 
For low water concentrations of the solution, the polymer is more likely 
to be found in the higher temperature chamber;
for high water concentrations the opposite is true.
Differences between theory and experiments are most pronounced at low
water concentrations, where 
our calculations overestimate the Soret effect.
This is a consequence of our choosing 
a mixed interaction parameter $\epsilon_\mathrm{ps}$ that emulates 
for short chains the poor solvent conditions that long PEO chains
experience in ethanol. 

Fig.~\ref{fig7} shows the composition differences $\delta c_i =
c_{i,\mathrm{A}}-c_{i,\mathrm{B}}$ between chambers A and B divided by
the temperature difference $\delta T$ 
for the three components (PEO, water, ethanol) as a function of 
water content of the solution. 
For PEO in a single solvent, ethanol or water, the composition difference 
$\delta c_\mathrm{p}$ of PEO is
balanced by the composition difference of the solvent.   
In general, the three composition differences add up to zero,  
$\delta c_\mathrm{p}+\delta c_\mathrm{w}+\delta c_\mathrm{s}=0$. 
The composition of the solvent is generally different in the hot and
cold chamber. For a solution with mass fraction of water
$c_\mathrm{w}\simeq 0.49$, for example, there is no difference in
ethanol content of the solution of the warm and cold chamber, while 
water is enriched in the cold chamber. 
Considering the ethanol content of the solvent, 
$\bar{c}_\mathrm{s}=c_\mathrm{s}/(c_\mathrm{s}+c_\mathrm{w})$, 
in each of the chambers, we find in our calculations that the solvent
in the warm chamber 
is always slightly richer in ethanol than the solvent in the cold
chamber. Furthermore, there is a range of water contents, 
approximately  10\% 
to 35\% by weight, where the difference in solvent composition
$\bar{c}_\mathrm{s,A}-\bar{c}_\mathrm{s,B}$ 
between the two chambers
is approximately constant at
$(\bar{c}_\mathrm{s,A}-\bar{c}_\mathrm{s,B})/\delta
T=2.5\times10^{-4}$~K$^{-1}$.

Our results for solvent composition reflect the
approximations in our calculations, in particular the random mixing
approximation for solvent-solvent interactions. When we perform
calculations on ethanol-water mixtures without polymer, we find a
Soret coefficient for water that is positive for all compositions and
decreases monotonously from 0.0069~K$^{-1}$ for ethanol-rich mixtures to 
0.0054~K$^{-1}$ for water rich mixtures. Experimental data 
\cite{ko88,zh96b,du02}, on the other hand, show the Soret coefficient
of water to change sign with increasing water content from around
0.006~K$^{-1}$ at 50~wt.\% to -0.008~K$^{-1}$ at 95~wt.\%

\begin{figure}
\vspace*{-0.3cm}
\includegraphics*[width=3.3in]{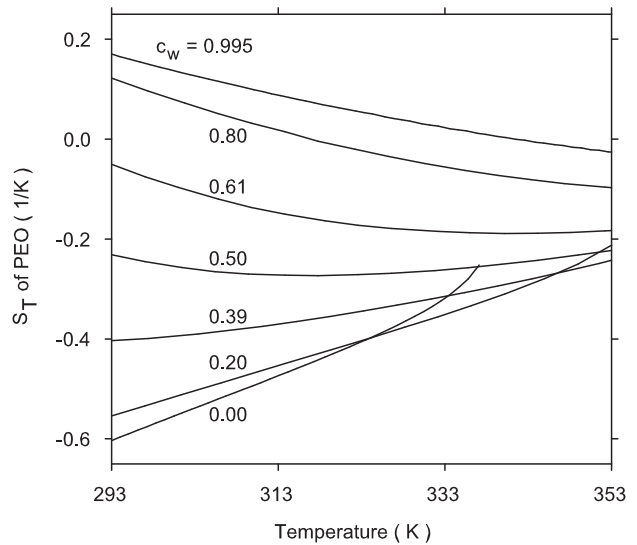}%
\caption{\label{fig8}
Soret coefficient of PEO in mixtures of ethanol and water at 
atmospheric pressure as a function of temperature for several
solvent compositions. The PEO concentration is
5g/L at 293~K. 
}
\end{figure}

In Fig.~\ref{fig8} we present results for the temperature
dependence of the Soret coefficient of PEO for several solvent
compositions. For high water concentration, the Soret coefficient is
positive at room temperature, decreases with increasing
temperature and changes sign at a temperature that depends on the water
content of the solution. For ethanol-rich solutions, the Soret
coefficient is negative at room temperature and increases with
increasing temperature. 
At the highest ethanol concentrations, an upturn of  
the $S_\mathrm{T}$ versus temperature curves is visible. This signals
the onset of evaporation, which occurs at increasingly higher
temperatures for solutions with larger water content. 
A comparison of Figs.~\ref{fig6} and Fig.~\ref{fig8} with the
chain-dimension graphs Fig.~\ref{fig4} (a) and (b) shows a correlation
between solvent quality and thermodiffusion. In general, as the
solvent quality increases, indicated by an increase in chain
dimensions, the Soret coefficient becomes more positive.

\section{Discussion}\label{discussion}

In this work, we have presented a two-chamber lattice-model approach
to determine Soret coefficients of polymer solutions. 
A dilute polymer solution is represented by a simple cubic lattice
occupied by a single polymer chain,
solvent particles and, in the case of a compressible solvent, voids. 
Exact enumeration results for an isolated chain allow us 
to construct partition functions for the polymer-solvent
system 
without invoking a random mixing approximation for contacts
with polymer sites.
Interactions between solvent particles, on the other hand, are
evaluated in a random-mixing approximation. 
Enumeration results are also used 
to calculate the radius of gyration of the chain in solution,  
which allows us to monitor the solvent quality of the solution.
In order to investigate thermodiffusion, 
we assume that the lattice is divided into two 
non-interacting sublattices of equal size  
that are maintained at slighty  different temperatures. 
For a given occupation of the
sublattices, the partition function of the combined system is a
product of the canonical partition functions of the sublattices. 
We consider all possible distributions of polymer chain and solvent
particles among the sublattices. 
The sum of states of the system is calculated by adding up the total
partition functions for all distributions, while average quantities
are calculated by performing the appropriate weighted sums. 
The Soret coefficient of the polymer is determined from 
the difference between the average polymer concentration in the
warm and cold chambers. 
As in the earlier work on heat of transport (cf.\ Denbigh \cite{de52}) 
kinetic energy contributions are neglected in our calculations. 
However, we do not approximate the heat of transport by a difference
in potential energy. Instead, the probability to find the polymer in
the warmer of the two chambers can be related to a difference in
average internal energy that reflects both enthalpic and entropic
contributions.

We considered first a lattice model for a dilute solution of a polymer 
in an incompressible, single-component solvent. 
In this case, the canonical partition function of a short chain 
($N_\mathrm{c}=17$) in
solution and the sum of states for the two-chamber system are
calculated exactly. We derived a simple relationship between the 
probability to find the polymer in the warmer chamber and the
internal-energy  
difference $U_\mathrm{nop}-U_\mathrm{pol}$, where 
$U_\mathrm{nop}$ represents the internal energy of a solvent filled
lattice and $U_\mathrm{pol}$ is the internal energy of a lattice
containing both solvent and polymer.
The resulting expression for the Soret coefficient shows 
how our approach relates to earlier work using the heat of transport
concept. It also illustrates that it is important to go beyond a
random mixing approximation when thermal diffusion in polymer
solutions is discussed. 
In agreement with experimental results for dilute 
solutions,\cite{ra02} the calculated Soret coefficients are
independent of concentration. 
Unfortunately, the short chains
considered in this work do not allow us to investigate the
experimentally 
observed\cite{ra02} scaling behavior of $S_\mathrm{T}$ with molecular
mass. 
We are currently working 
on a simulation approach that allows us to treat longer
chains\cite{lu03} and will investigate the scaling laws in the near
future. 
Our results presented in Fig.~\ref{fig3} show that both positive and
negative Soret coefficients are realized in the incompressible case
and that the sign of $S_\mathrm{T}$ need not be the same for all
temperatures. A comparison with 
Fig.~\ref{fig1} reveals a correlation between 
Soret coefficients and chain dimensions for two of the three cases
presented; the probability to find the
polymer in the cooler chamber increases as the solvent quality
increases. This trend appears to be typical and, as discussed below,
may provide an explanation why 
negative Soret coefficients for polymer solutions are so rarely observed.

In order to describe PEO chains in a mixed solvent of ethanol and
water, we extended 
the simple lattice model by introducing two types of solvent
particles and adding voids to account for compressibility.
Hydrogen bonding between PEO and water molecules is modeled through 
an orientational degree of freedom of the water particles.
One of the six faces of an elementary cube representing a water particle
has strongly attractive interactions with PEO sites, while the
PEO interactions of the remaining five faces are the same as those
between ethanol and PEO. 
The lattice model for dilute solutions of PEO in ethanol/water mixtures
has eleven system-dependent parameters that are determined from static
properties as described in Appendix \ref{appa} and summarized in Table 
\ref{tabsys}. With these parameters, the model reproduces some of the
important thermodynamic properties of the system. In particular, 
for PEO in mixed solvents, the solvent quality as monitored by the
radius of gyration increases as the water content of the solution
increases\cite{de02,de02b}. Similarly, increasing the temperature
increases the solvent 
quality for mixtures with low water content. For mixtures with high
water content, on the other hand, increasing the temperature reduces
the solvent quality in agreement with observations on PEO in water
(cf.\ Ref. \onlinecite{do02}). 

Our two-chamber approach allows us to calculate Soret coefficients of
PEO for given temperature, pressure and composition of the solvent. 
As expected for dilute solutions,\cite{ra02} the results are
independent of the polymer concentration. In qualitative agreement
with  experimental data of De Gans {\it et al.}\cite{de02,de02b} and
Kita {\it et al.}\cite{ki03}, the calculated Soret coefficients are 
negative for solutions with low water content and positive for
solutions with high water content. 
While we do not expect quantitative agreement between our simple lattice
model calculations and experimental data, we would like to discuss
some future work that may help us gain a better understanding of
thermodiffusion in this system.
The evaluation of Soret coefficients with the aid of Eq.~(\ref{st}) is
valid only when pressure differences between the chambers can be
neglected. We observe very small differences between the
pressures in the two chambers and do not expect them to contribute
significantly, but we are investigating means to 
correct for this effect.
Interactions between different polymer chains are not considered in
our calculations, which apply to dilute solutions.  
However, the polymer solutions used in the
experiments\cite{de02,de02b,ki03} were probably in the semi-dilute
regime. In typical semi-dilute solutions, the magnitude of the Soret
coefficient decreases with increasing concentration.\cite{ra02} 
It will be interesting to investigate the effect of chain-chain
interactions on negative thermodiffusion coefficients. 
Significant thermal diffusion of the solvent particles is an issue that
complicates the comparison between theory and experiment 
for intermediate water concentrations.\cite{ki03} 
Eq.~(\ref{ternst}) for the Soret effect
in ternary mixtures suggests that several diffusion coefficients need
to be considered carefully for an experimental determination of the
Soret coefficient in this regime.
Hydrogen bonding between solvent molecules may also affect the thermal
diffusion of the polymer. 
Our current calculations account for specific interactions only
between PEO and water molecules.  
We will extend our model to account for hydrogen bonds between solvent
molecules and investigate how it affects the Soret coefficient of the
polymer.

Finally, we have also investigated the temperature dependence of the
Soret coefficient of PEO and found that a change in temperature may
induce a change in sign of $S_\mathrm{T}$. 
A comparison of the results for the radius of gyration of the chain
and the values of the Soret coefficients reveals a close relationship
between the solvent quality and the partitioning of the polymer
between the chambers. 
In general, the Soret coefficient of the polymer becomes more
positive as the solvent quality increases. 
A typical experiment on polymers in good solvents is thus expected
to yield positive Soret coefficients. 
We expect negative Soret coefficients to be observed for polymers 
that would be insoluble were it not for specific interactions between
solvent molecules and sites on the polymer. 
It appears that both polymer systems for which negative Soret
coefficients have been observed, 
the solutions of PEO in a mixed ethanol/water
solvent\cite{de02,de02b,ki03} and the solution of poly(vinyl alcohol) in
water,\cite{gi77,ea82} belong to this category.

\begin{acknowledgments}
We would like to thank S. Wiegand and M.~P. Taylor for many helpful
discussions and S. Wiegand and her coworkers for making experimental
data available to us prior to publication. 
Financial support through the National Science Foundation (NSF
DMR-0103704), the Research Corporation (CC5228), and the Petroleum
Research Fund (PRF \#364559 GB7) is gratefully acknowledged.
\end{acknowledgments}

\appendix

\section{Determination of system-dependent parameters for PEO in
ethanol water mixtures}\label{parameters}\label{appa}

Our lattice model for PEO in ethanol/water
mixtures has eleven system-dependent parameters.
Seven of these parameters 
describe the thermodynamics of the pure components 
while the remaining four are needed for the description of the
mixtures. 

For each of the pure components (ethanol, water, PEO) we estimate 
interaction energies $\epsilon_{ii}$ and the volume $v_0$ per lattice
site 
from a comparison of tabulated values \cite{CRC,va72}
for the density and the thermal expansion coefficient at standard
temperature and pressure with values calculated from lattice-fluid
equations of state \cite{la00,sa76,co81}, which treat nearest neighbor
contacts in random mixing approximation. 
For water and ethanol we employ a simple lattice fluid model, 
the Sanchez and Lacombe equation-of-state applied to
monomers,\cite{la00,sa76}
\begin{equation}\label{eos}
P=\frac{1}{2v_0}z\epsilon\phi^2-\frac{RT}{v_0}\ln(1-\phi) ,
\end{equation}
where $R$ is the ideal gas constant and 
$v_0$ is the volume per mole of lattice sites. 
$\phi$ is the filling fraction of the lattice and related in this work
to the mass density $\rho$ by 
\begin{equation}\label{phirho}
\phi=\frac{sv_0\rho}{M} ,
\end{equation}
where $M$ is the molecular mass of the component and $s$ is a scale
factor that is discussed below.
Since the tabulated values for the properties of PEO\cite{va72}
correspond to large molecular mass, we use an equation of state by
Costas and Sanctuary\cite{co81} in the limit of infinite chain length 
\begin{eqnarray}
P&=&\frac{1}{2v_0}z\epsilon
\left( \frac{(z-2)\phi}{z-2\phi} \right)^2
-\frac{RT}{v_0}\ln(1-\phi) \nonumber \\
& & \mbox{}
+\frac{zRT}{2v_0}\ln\left(1-\frac{2}{z}\phi\right)
\end{eqnarray}

Comparing calculated and tabulated values of thermodynamic properties
for each substance individually, we arrive at three different values
for the volume $v_0$. For a description on a common lattice, we chose
a value for $v_0$ and scale factors $s_i$ that allow us to reproduce 
the tabulated densities at standard conditions and obtain
reasonable estimates for the thermal expansion coefficients. 
The resulting values are presented in Table~\ref{tabsys}. The scale
factor for ethanol is chosen as $s_\mathrm{s}=1$ and implies that 
each ethanol molecule occupies one lattice site. The scale factors for
water and PEO are smaller than unity.
For water,
there are $1/s_\mathrm{w}\approx 3$ 
molecules associated with each site. 
For PEO, our short chains ($N_\mathrm{c}=17$) correspond to 
physical chains with 
$N_\mathrm{c}/s_\mathrm{p}\approx 27$ repeat units. Since each repeat
unit has a molecular mass of $M_\mathrm{mono}=44.1$~g/mol, the molecular
mass of our chains is
$M_\mathrm{p}=N_\mathrm{c}M_\mathrm{mono}/s_\mathrm{p}\approx 1187$~g/mol.

\begin{table}
\caption{System-dependent parameters for the \\ PEO/ethanol/water
system
}\label{tabsys}
\begin{ruledtabular}
\begin{tabular}{l|l|l}
\multicolumn{3}{l}{ lattice site volume 
 $v_0=5.255\times10^{-5}$m$^3$/mol }  \\ 
& $\epsilon_{ij}$ in J/mol & scale factors \\
ethanol & $\epsilon_\mathrm{ss}=-2306$ & $s_\mathrm{s}=1$ \\
water & $\epsilon_\mathrm{ww}=-3306$ & $s_\mathrm{w}=0.3362$ \\
PEO & $\epsilon_\mathrm{pp}=-1153$ & $s_\mathrm{p}=0.6318$  \\
& mixed interactions & \\
ethanol/water & 
$\epsilon_\mathrm{ws}=-\sqrt{\epsilon_\mathrm{ww}\epsilon_\mathrm{ss}}$
& \\
PEO/water & $\epsilon_\mathrm{pw,n}=2660$ & \\
 & $\epsilon_\mathrm{pw,s}=-8020$ & \\
PEO/ethanol &  $\epsilon_\mathrm{ps}=2660$  & \\
\end{tabular}
\end{ruledtabular}
\end{table}

\begin{table}
\caption{Comparison of tabulated\protect{\cite{CRC,va72}} and 
calculated values for the density $\rho$ and thermal expansion
coefficient $\alpha$ at a temperature of 293~K and a pressure of 
0.1~MPa}\label{compare}
\begin{ruledtabular}
\begin{tabular}{l|l|l|l|l}
&
\multicolumn{2}{c|}{$\rho$ (kg/m$^3$)} &
\multicolumn{2}{c}{$\alpha$ (K$^{-1}$)} \\
& tabulated & calculated & tabulated & calculated \\
ethanol & 789 & 788.8 & $1.43\times10^{-3}$ & $1.79\times10^{-3}$ \\
water & 998.2 & 998.5 & $2.10\times10^{-4}$ & $3.24\times10^{-4}$ \\
PEO & 1125 & 1125.0 & $7.2\times10^{-4}$ & $3.83\times10^{-4}$ 
\end{tabular}
\end{ruledtabular}
\end{table}

The only additional parameters required for the description of 
mixtures of the three substances are the mixed interaction energies. 
Since we are using a very simple model to describe the solvent, 
we employ Berthelot's geometric mean combining rule
$\epsilon_\mathrm{ws}=-\sqrt{\epsilon_\mathrm{ww}\epsilon_\mathrm{ss}}$
for interactions between ethanol and water.
This approximation leads to deviations between calculated and
tabulated~\cite{CRC} values for the density of less than 2\% for all
compositions at a temperature of 293~K and a pressure of 0.1~MPa. 
A better description of ethanol and water can be achieved when
specific interactions between the molecules are included. 
We are currently working on such an extension of our model.

With $\epsilon_\mathrm{ss}$ and $\epsilon_\mathrm{pp}$ fixed, the
mixed interaction energy $\epsilon_\mathrm{ps}$ determines the
solubility of the polymer in ethanol. The miscibility of polymers is
strongly dependent on chain length. 
Unfortunately, solubility data for the short chains of our
calculations ($M_\mathrm{p}\approx 1187$~g/mol ) are not available. 
The long PEO chains ($M\approx 236,000$~g/mol) 
employed in the experiments \cite{de02} are insoluble in ethanol at
room temperature and become soluble only for temperatures above 
323~K. We obtained an estimate for $\epsilon_\mathrm{ps}$ 
for long PEO
chains with the aid of the Born-Green-Yvon lattice model of 
Lipson and coworkers \cite{li93,lu98,lu98d}. Employing the pure
component parameters determined above, we adjusted $\epsilon_\mathrm{ps}$ 
until the calculated phase transition temperature was near 323~K.
The resulting value is 
$\epsilon_\mathrm{ps}=-1141$~J/mol. 
This value makes ethanol an excellent solvent for the
short chains, in agreement with experimental observations, but, 
unfortunately, it does not allow us to use our short-chain
calculation to explore the effect of changing 
solvent quality with temperature and solvent composition. 
In order to find a more suitable interaction parameter, 
we use the radius of gyration squared,
$R_\mathrm{g}^2$, as an indicator for solvent quality. 
We estimate a value of $\epsilon_\mathrm{ps}$ from the condition that
the calculated value of $R_\mathrm{g}^2$ 
at a temperature of 323~K is
equal to $R_\mathrm{g}^2(\theta^*)=4.13$, 
the chain dimensions of an
isolated chain at the $\theta$ temperature of the infinite chain.
This value for $\epsilon_\mathrm{ps}$ is included in
Table~\ref{tabsys} and used in our calculations.
Fig.~\ref{fig4} (b) includes
a graph of calculated chain dimensions as a function of temperature
for PEO in ethanol. A comparison with Fig.~\ref{fig1} shows that the
chains are not collapsed at room temperature, indicating that the
short chains are soluble in ethanol.
However, as in the case of the long chains, 
solvent quality improves with
increasing temperature over a large temperature range. 

The interactions between PEO and water are described by two
interaction parameters. Following Jeppesen and Kremer \cite{je96} the
non-specific interactions $\epsilon_\mathrm{pw;n}$ 
are assumed to correspond to bad-solvent
conditions and are approximated here by the value for the ethanol-PEO
interactions. The specific (hydrogen-bonding) interactions 
$\epsilon_\mathrm{pw;s}$  are strongly attractive. As the remaining
free parameter, they determine the solubility of PEO in mixtures of 
ethanol and water. Experiments suggest~\cite{de02} that the transition
between poor and good solvent conditions occurs for mass fractions of
water between 5\% and 10\%. 
We estimate a value for $\epsilon_\mathrm{pw;s}$ from the condition
that the calculated value of $R_\mathrm{g}^2$ in a mixture with
$c_\mathrm{w}=0.1$ is equal to $R_\mathrm{g}^2(\theta^*)=4.13$.

\section{Determination of lattice size and occupation numbers}\label{appb}

In order to calculate properties of PEO solutions at given 
temperature, pressure, and composition, the corresponding lattice size
$N$ and occupation numbers $N_i$, $i\in\{p,s,w\}$ have to be determined. 
In the results presented here, we start at room temperature, $T=293$~K,
and atmospheric pressure, $P\approx 0.1$~MPa, and specify the PEO
concentration as mass per unit volume. 
This sets the lattice size, since the lattice contains a single polymer
chain occupying $N_\mathrm{c}=17$ lattice sites
\begin{equation}\label{Ncalc}
N=\frac{M_\mathrm{p}}{v_0}/\rho_\mathrm{p} ,
\end{equation}
where 
$M_\mathrm{p}=N_\mathrm{c}M_\mathrm{mono}/s_\mathrm{p}\approx 1187$~g/mol
is the molecular mass of the chains and 
$\rho_\mathrm{p}$ is the PEO concentration in g/L.
If two solvents are present, we specify the occupation number for one
them, say water ($N_\mathrm{w}$). 
The remaining $N-N_\mathrm{c}-N_\mathrm{w}$ lattice
sites are then divided between ethanol ($N_\mathrm{s}$) and voids
($N_\mathrm{v}$) so that the pressure
calculated according to Eq.~(\ref{pres}) is approximately equal to
atmospheric pressure. Since the lattice occupation numbers are
integers, the targeted pressure value cannot be reached exactly. 
We use an iterative algorithm that identifies occupation numbers
$N_\mathrm{s}$ and $N_\mathrm{v}$ corresponding to the closest
calculated pressure above the target pressure.
The uncertainty introduced into our results by the integer nature of
the occupation numbers decreases as the size of the lattice
increases.  
Since the number of lattice
sites is set by the PEO concentration, and since our results for the
Soret coefficient of PEO are independent of its concentration, 
large lattices can be used to reduce the uncertainty in the calculated
values. Unfortunately, the
computation time grows rapidly with increasing lattice size, so that a
compromise has to be found. 
In this work, we use $\rho_\mathrm{p}=5$~g/L at $T=293$~K 
and target pressure $P=0.1$~MPa. This corresponds to 
a lattice size of  $N=4516$ sites and yields
calculated pressure values from 0.13~MPa for ethanol rich mixtures to 
0.4~MPa for water rich mixtures. 
In order to perform calculations at elevated temperature or pressure,
we retain the occupation numbers for the filled sites, $N_\mathrm{c}$, 
$N_\mathrm{w}$, and $N_\mathrm{s}$, but change the total number of
lattice sites by adding or removing voids to reach the desired
conditions.

The concentration of each of the components in the system is
related to the lattice occupation through
\begin{equation}\label{concentration}
\rho_i=\frac{m_i}{Nv_0}\:, \mbox{\hspace{1em}} 
\sum_i\rho_i=\rho\:, \mbox{\hspace{1em}} 
i\in\{s,w,p\} 
\end{equation}
where
\begin{equation}\label{masses}
m_\mathrm{s}=\frac{N_\mathrm{s}M_\mathrm{s}}{s_\mathrm{s}} , \:
m_\mathrm{w}=\frac{N_\mathrm{w}M_\mathrm{w}}{s_\mathrm{w}} , \:
m_\mathrm{p}=\frac{N_\mathrm{c}M_\mathrm{mono}}{s_\mathrm{p}} , 
\end{equation}
and $M_\mathrm{s}=46.07$~g/mol and $M_\mathrm{w}=18.0$~g/mol are the
molecular masses of ethanol and water, respectively.
The mass fractions $c_i$ of the components are given by
\begin{equation}\label{massfrac}
c_i=\rho_i/\rho .
\end{equation}

\bibliography{dart}

\end{document}